\documentclass[preprint, aps, prd, preprintnumbers, superscriptaddress, nofootinbib, floatfix]{revtex4-1}

\usepackage{amsmath}
\usepackage{bm}
\usepackage{amsfonts}
\usepackage{graphicx}
\usepackage{hyperref}
\usepackage{color}
\usepackage{ulem}

\hypersetup{
colorlinks=true,
linkcolor=blue,
linktoc=page,
citecolor=blue,
urlcolor=blue}

\begin{document}

\title{A bound on thermal $y$-distortion of the cosmic neutrino background}

\newcommand{\FIRSTAFF}{\affiliation{Department of Physics, University at Buffalo, Buffalo, NY 14260, USA}}
\newcommand{\SECONDAFF}{\affiliation{Departament de Física Teórica and IFIC, Universitat de València-CSIC, E-46100, Burjassot, Spain}}

\author{Gabriela Barenboim}
\email[Electronic address: ]{gabriela.barenboim@uv.es}
\SECONDAFF
\author{Héctor Sanchis}
\email[Electronic address: ]{hector.sanchis@ific.uv.es}
\SECONDAFF
\author{William H. Kinney}
\email[Electronic address: ]{whkinney@buffalo.edu}
\FIRSTAFF
\author{Diego Rios}
\email[Electronic address: ]{diegorio@buffalo.edu}
\FIRSTAFF

\date{\today}
\begin{abstract}
We consider the possibility that the cosmic neutrino background might have a nonthermal spectrum, and investigate its effect on cosmological parameters relative to standard $\Lambda$-Cold Dark Matter ($\Lambda$CDM) cosmology. As a specific model, we consider a thermal $y$-distortion, which alters the distribution function of the neutrino background by depleting the population of low-energy neutrinos and enhancing the high-energy tail. We constrain the thermal $y$-parameter of the cosmic neutrino background using Cosmic Microwave Background (CMB) and Baryon Acoustic Oscillation (BAO) measurements, and place a $95\%$-confidence upper bound of $y \leq 0.043$. The $y$-parameter increases the number of effective relativistic degrees of freedom, reducing the sound horizon radius and increasing the best-fit value for the Hubble constant $H_0$. We obtain an upper bound on the Hubble constant of $H_0 = 71.12\ \mathrm{km/s/Mpc}$ at $95\%$ confidence, substantially reducing the tension between CMB/BAO constraints and direct measurement of the expansion rate from Type-Ia supernovae. Including a spectral distortion also allows for a higher value of the spectral index of scalar fluctuations, with a best-fit of $n_{\mathrm{S}} = 0.9720 \pm 0.0063$, and a $95\%$-confidence upper bound of $n_{\mathrm{S}} \leq 0.9842$. 
\end{abstract}

\maketitle

\section{Introduction}
The standard $\Lambda$-Cold Dark Matter ($\Lambda$CDM) cosmological model is a remarkably successful description of the observed universe, from the epoch of recombination and the formation of the Cosmic Microwave Background (CMB) through the era of structure formation in the universe, and up to the present epoch of dark energy domination and accelerated expansion. The model fits cosmological observations at a variety of scales and redshifts to a simple set of six parameters, describing a nearly scale-invariant primordial spectrum of density perturbations, the densities of baryons and dark matter, the angular diameter distance to the surface of last scattering, and the redshift of star formation and the subsequent reionization of the universe. While this simple benchmark model is a good fit to cosmological observations, there are a few tensions affecting cosmological parameters. The most significant of these is the disagreement between direct measurement of the Hubble Constant $H_0$ using standard candles such as Type-Ia supernovae resulting in a value of $72.3 \pm 1.4\ \mathrm{(stat)}\ \pm 1.4\ \mathrm{(syst)}\ \mathrm{km\ s^{-1}\ Mpc^{-1}}$ \cite{Riess:2016jrr,Galbany:2022zir}, and measurement of $H_0$ obtained from measurements of the anisotropy of the Cosmic Microwave Background, which give a value of $67.4 \pm 0.5 \ \mathrm{km\ s^{-1}\ Mpc^{-1}}$ \cite{Planck:2018vyg}, resulting in a tension in excess of $4 \sigma$. The explanation for this tension is currently unknown. 

An important physical feature of the data is that standard candle measurements directly measure the expansion rate $H_0$, while constraints from the CMB are \textit{indirect}, and rely on the $\Lambda$CDM model as an underlying assumption. This opens the possibility that the apparent tension in the Hubble parameter could be an artifact of assuming an incorrect or incomplete cosmological model, and could be explained by an extension to $\Lambda$CDM involving new physics. This is easier said than done \cite{Cai:2021weh}, and many such extensions have been proposed, with none providing a compelling resolution (for reviews see Refs. \cite{Verde:2019ivm,Knox:2019rjx,DiValentino:2021izs}). A particularly promising candidate for new physics is the neutrino sector; existing proposals for mitigating the Hubble tension include interaction between neutrinos and dark energy prior to recombination \cite{Sakstein:2019fmf} (see, however, \cite{deSouza:2023sqp,CarrilloGonzalez:2023lma}), and a lepton asymmetry in the neutrino sector \cite{Barenboim:2016lxv}. It has recently been noted that current constraints actually favor a \textit{negative} value for the sum of neutrino masses, suggesting the influence of new physics \cite{eBOSS:2020yzd,Craig:2024tky}.

In this paper we consider the possibility of a non-thermal spectrum for the cosmic neutrino background. In the standard cosmological model, there are two backgrounds of relic radiation left over from the early hot Big Bang universe: photons from the epoch of recombination (the CMB), and neutrinos, which decoupled shortly before the epoch of $e^{\pm}$ annihilation. The thermal spectrum of the CMB has been measured to have a a temperature of $2.7\ \mathrm{K}$, and exhibits an exactly thermal spectrum, measured with extremely high precision by the FIRAS instrument \cite{Mather:1993ij}. The cosmic neutrino background, by contrast, has only been detected indirectly, and an underlying assumption of the $\Lambda$CDM cosmological model is that the neutrino background is, like the CMB, thermal, with a temperature of $1.95\ \mathrm{K}$. In this paper, we investigate the consequences of relaxing the assumption that the cosmic neutrino background has an exactly thermal spectrum. This has been considered previously in the literature for modifications such as the addition of a Gaussian on top of the thermal distribution \cite{Cuoco:2005qr,Alvey:2021sji}, leaving the energy density fixed, but changing the neutrino number density, and by an effective chemical potential in the neutrino sector \cite{Kinney:1999pd,Kawasaki:2000en,Barenboim:2016lxv}.  Here, we consider a $y$-type distortion of the neutrino distribution function, which distorts the occupation function while leaving the number density constant, adding one parameter to the standard $\Lambda$CDM model which behaves similarly to changing the overall number of relativistic degrees of freedom $N_{\mathrm{eff}}$. Our main result is an upper bound on the thermal $y$-parameter of the cosmic neutrino background of $y \leq 0.043$ at $95\%$ confidence. We find that the $y$-parameter is strongly correlated with both the sound horizon radius and the Hubble constant, and substantially broadens the allowed region for the Hubble constant from the CMB and Baryon Acoustic Oscillation (BAO) data. The paper is organized as follows: Sec. \ref{sec:yDistortion} describes the physics of the spectral distortion, \ref{sec:Constraints} shows constraints from CMB and BAO, and \ref{sec:Conclusions} presents a summary and conclusions. 

\section{Thermal Distortions in the Neutrino Sector}
\label{sec:yDistortion}
A  $y$-distortion refers to a specific type of distortion in the spectrum of the CMB  radiation, caused by inverse Compton scattering of CMB photons off hot electrons, leading to a characteristic change in the distribution of photon energies. The best-known example of a $y$-distortion is the thermal Sunyaev--Zel'dovich effect \cite{1980ARA&A..18..537S}, in which CMB photons passing through a cluster of galaxies encounter hot electrons in the intracluster medium. These electrons are much hotter than the CMB photons, typically with temperatures in the range of millions of Kelvin, but are optically thin, resulting in single-scattering events which selectively up-scatter low-energy CMB photons to higher energies, resulting in a characteristic distortion of the underlying black-body spectrum. 

The $y$  parameter quantifies the amount of energy transferred from the hot electrons to the CMB photons and it is defined as 
\begin{equation}
y = \int{\left( \frac{k_B T_\mathrm{e}}{m_\mathrm{e} c^2} \right) N_\mathrm{e} \sigma_\mathrm{T} d\ell},
\end{equation}
where $k_\mathrm{B}$  is the Boltzmann constant, $  T_\mathrm{e}$ is the electron temperature, $m_\mathrm{e}$  is the electron mass, $N_\mathrm{e}$  is the electron number density, $ \sigma_\mathrm{T} $ is the Thomson scattering cross-section, and $d \ell$  is the path length through the electron gas. The $y$-distortion alters the CMB spectrum from a perfect blackbody by introducing a decrement in intensity at lower frequencies and an increment at higher frequencies, of the form
\begin{equation}
  f_\gamma^y(x)= f_\gamma(x) + \delta f_\gamma (x) = \frac{1}{e^x-1}  \left[1+ y \frac{e^x x}{e^x-1} \left(x\frac{e^x+1}{e^x-1}-4\right)\right]
\end{equation}
where $f_\gamma(x) = (e^x -1)^{-1}$. The total number of photon remains constant 
\begin{equation}
  N \propto \int x^2  \;  f_\gamma^y (x) \; dx =   \int x^2  \;  f_\gamma  (x) \; dx, 
\end{equation}
and the energy injection is given by 
\begin{equation}
  \delta \rho_\gamma  = \int x^3 \; \delta f_\gamma^y (x) \; dx = 4 \, y \; \rho_\gamma  
\end{equation}
where $\rho_\gamma $ is the energy contained in the undistorted spectrum.

\begin{figure}[!h]
\centering
\begin{minipage}{.5\textwidth}
\includegraphics[width=8.cm]{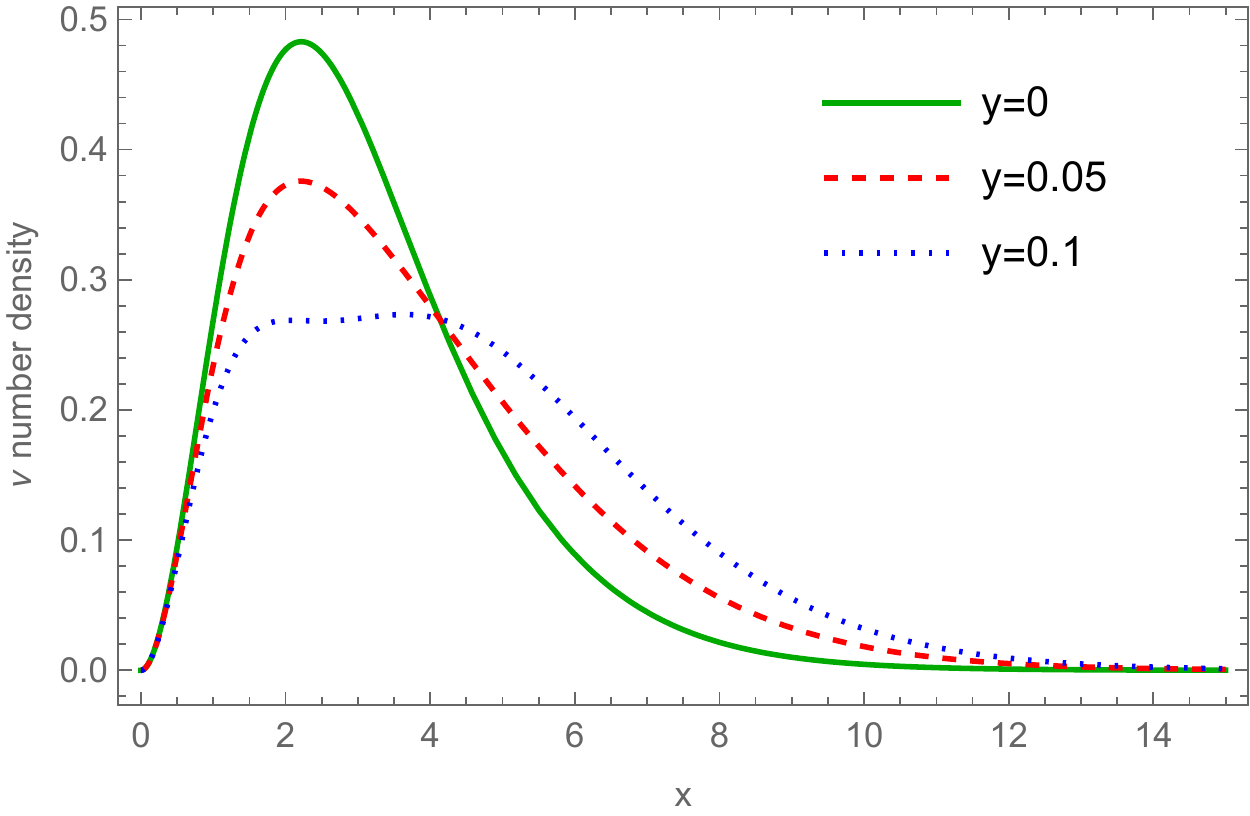}
\end{minipage}%
\begin{minipage}{.5\textwidth}
\includegraphics[width=8.cm]{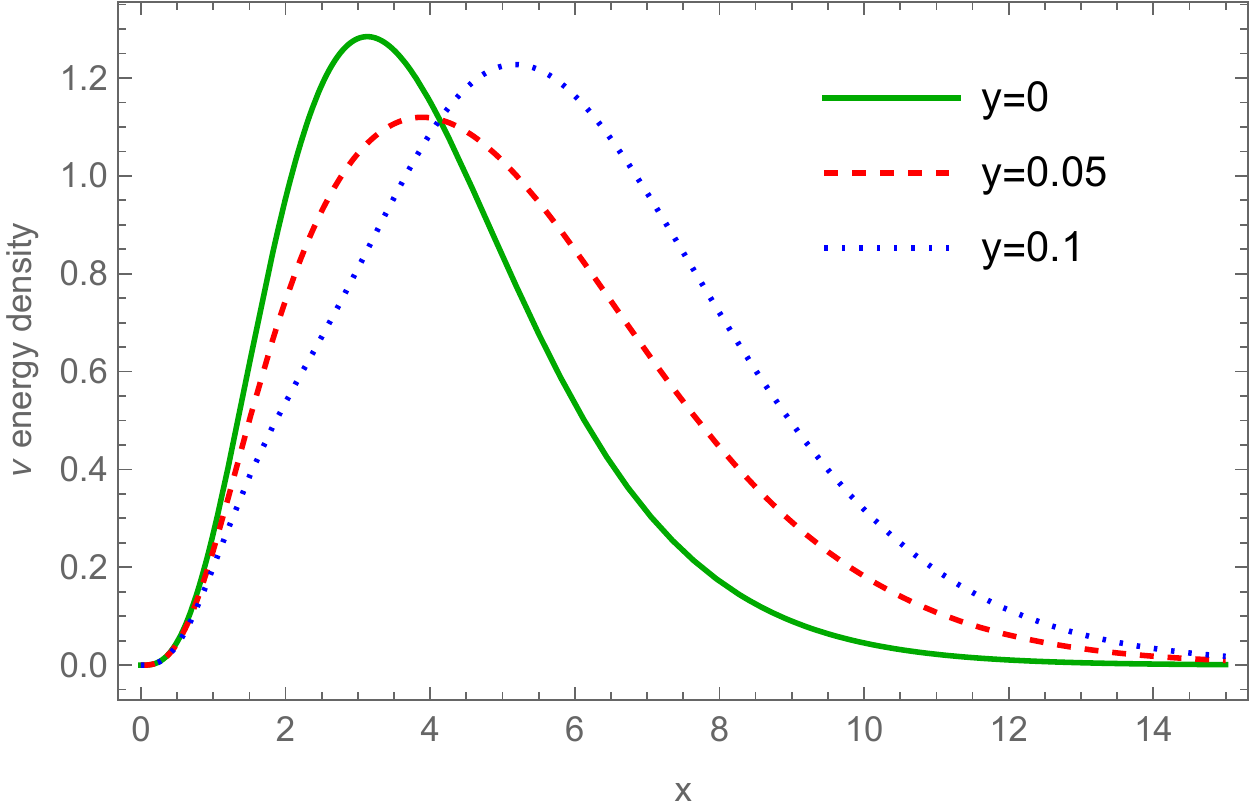}
\end{minipage}
\caption{Illustration of the effect of a $y$-distortion on the neutrino number density (left), and the energy density (right), for $y = 0$ (green, solid), $y = 0.05$ (red, dashed), and $y = 0.1$, (blue, dotted).}
\label{fig:ydistortions}
\end{figure}

In this paper, we consider the possibility that a similar non-thermal distortion could apply to the cosmic neutrino background; we propose no specific mechanism, but instead use a $y$-distortion as a specific one-parameter phenomenological description, valid as long as the physics generating the distortion is dominated by single-scattering events which preferentially up-scatter low-energy neutrinos to higher energy. Accounting for  the difference between bosonic and fermionic distribution functions, we assume a spectral distortion of the form,
\begin{equation}
  f_\nu^y(x)= f_\nu(x) + \delta f_\gamma (x) = \frac{1}{e^x+1}  \left[1+ y \frac{e^x x}{e^x+1} \left(x\frac{e^x-1}{e^x+1}-4\right)\right]
\end{equation}
As in the case of photons, this distortion does not change the total number of neutrinos but increases their total energy (Fig. \ref{fig:ydistortions}). 

Thermal distortion of the neutrino distribution will have several effects on cosmology. First, the spectral distortion induces a shift in the energy density of the neutrino background, 
\begin{equation}
    \frac{\delta \rho_\nu}{\rho_\nu} = 4 y,
\end{equation}
This will consequently add to the total radiation density,
\begin{equation}
    \rho_{\mathrm{R}} = \rho_\gamma + \rho_\nu = g_* \rho_\gamma,
\end{equation}
where
\begin{equation}
    g_* = 2 + 2 N_{\mathrm{eff}} \left(\frac{7}{8}\right) \left(\frac{4}{11}\right)^{4/3}
\end{equation}
Note that this expression, strictly speaking, only applies in thermal equilibrium. However, for a small $y$-distortion, we can arrive at a qualitative understanding of the physical effects by approximating the spectral distortion as a shift in effective thermal-equilibrium quantities. First, matter/radiation equality will occur slightly later compared to the fiducial $\Lambda$CDM cosmology. Taking $z_{\mathrm{eq}}$ the redshift of matter/radiation equality in the absence of a $y$-distortion, the redshift $\tilde z_{\mathrm{eq}}$ including the spectral distortion will be approximately
\begin{equation}
    \frac{1 + z_{\mathrm{eq}}}{1 + \tilde z_{\mathrm{eq}}} = 1 + 2 \left(\frac{\left(7/8\right) \left(4/11\right)^{4/3} N_{\mathrm{eff}}}{1 + \left(7/8\right) \left(4/11\right)^{4/3} N_{\mathrm{eff}}}\right) y \simeq 1 + 0.5 y,
\end{equation}
for $N_{\mathrm{eff}} = 3.046$. Second, the spectral distortion has the effect of allowing the expansion rate prior to last scattering to be slightly larger at the same redshift due to the injection of energy into relativistic degrees of freedom. Since matter/radiation equality occurs later, and the expansion rate is increased, the sound horizon at last scattering $r^\star_\mathrm{s}$ will be smaller, \cite{Knox:2019rjx}
\begin{equation}
    r^\star_\mathrm{s} = \int_0^{t_\star}{\frac{c_\mathrm{S}\left(t\right) dt}{a\left(t\right)}} = \int_{z_\star}^\infty{\frac{c_\mathrm{S}\left(z\right) dz}{H\left(z\right)}},
\end{equation}
where $c_\mathrm{S}$ is the sound speed, and $z_\star$ is the redshift of last scattering. This decrease in the sound horizon at last scattering affects the constraint on the Hubble parameter, since the angular diameter of the sound horizon $\theta_\star$ is fixed by observation, but
\begin{equation}
    \theta_\star = \frac{r_{\mathrm{s}}^\star}{D_A^\star},
\end{equation}
where $D_A^\star$ is the angular diameter distance to the surface of last scattering,
\begin{equation}
    D_A^\star = \int_0^{z_\star}{\frac{dz}{H\left(z\right)}},
\end{equation}
must also decrease in order to keep $\theta_\star$ fixed, which means we must \textit{increase} the expansion rate at late time. In this sense, the $y$-distortion behaves similarly to an increase to $\delta N_{\mathrm{eff}} / N_{\mathrm{eff}} \sim 4 y$, since the entire neutrino spectrum is relativistic, and the spectral distortion increases the total energy in the neutrino background. Models including varying $N_{\mathrm{eff}}$ have been well-studied in the literature \cite{Knox:2019rjx,2013PhRvD..87h3008H}. We expect that a positive $y$-parameter will \textit{decrease} $r_{\mathrm{s}}^\star$ and the related acoustic drag scale $r_\mathrm{s}^{\mathrm{drag}}$, and \textit{increase} the best-fit Hubble parameter $H_0$.

The $y$-distortion in the neutrino background in principle differs from simply varying $N_{\mathrm{eff}}$ in two key ways. The first is that the $y$-distortion may exist before neutrino decoupling, or be induced by processes that occur afterward. That is, if we assume entirely standard nucleosynthesis, with $N_{\mathrm{eff}} = 3.05$, and a Helium abundance of $Y_{\mathrm{P}} = 2.4709$, consistent with the bound from Ref. \cite{Pitrou:2018cgg} of $Y_{\mathrm{P}} = 0.24709 \pm 0.00017$, this (slightly) differs from simply varying $N_{\mathrm{eff}}$, because $N_{\mathrm{eff}}$ during nucleosynthesis is correlated with the Helium abundance, such that a larger $N_{\mathrm{eff}}$ also requires a higher value of $Y_{\mathrm{P}}$. The Helium abundance influences CMB observables primarily through its effect on the electron abundance, and therefore the damping tail of the CMB power spectrum \cite{2013PhRvD..87h3008H}, which for the parameter ranges considered here can be as large as $2\%$. In practice, we find that has no discernible effect on the bounds on other parameters, and in what follows, we enforce the standard BBN constraint between $N_{\mathrm{eff}}$ and $Y_{\mathrm{P}}$.

A second difference is that neutrinos are massive; introducing a $y$-distortion slightly depletes the population of low-energy neutrinos and enhances the population of high-energy neutrinos. Since it is the low-energy tail of the neutrino spectrum which becomes nonrelativistic at late times, we expect a small decrease in the neutrino contribution to Cold Dark Matter (CDM) during structure formation. In practice, for the time being, this latter effect is negligible, and the effect of the $y$-distortion is dominated by its contribution to the relativistic background. Future measurements of the large scale structures may allow the discrimination between the two scenarios. However, one should keep in mind that $N_{\mathrm{eff}}$ encompasses a variety of different models, from sterile massive neutrinos to extra massless degrees of freedom, each one having a different large scale structure signature. The sign of the $y$-parameter can also in principle be negative, \textit{i.e.} optically thin downscattering of neutrinos from high energy to low energy. A $y$-distortion therefore offers the possibility to explore scenarios with a reduction of $N_{\mathrm{eff}}$, something that can not be achieved by adding extra relativistic degrees of freedom. A negative $y$-distortion could also serve to compensate for the addition of extra relativistic species.

In the next section, we discuss parameter constraints. 

\section{Parameter Constraints}
\label{sec:Constraints}

We consider a fiducial $\Lambda$CDM model consisting of the following six parameters:
\begin{itemize}
    \item{Baryon density $\Omega_{\mathrm{b}} h^2$}
    \item{Cold Dark Matter density $\Omega_{\mathrm{C}} h^2$}
    \item{Angular diameter of the sound horizon at last scattering $\theta_\star$}
    \item{Reionization optical depth $\tau$}
    \item{Amplitude of scalar density perturbations $A_\mathrm{S}$}
    \item{Spectral index of scalar density perturbations $n_\mathrm{S}$}
\end{itemize}
We assume a flat universe, $\Omega_\mathrm{tot} = 1$, no primordial tensor fluctuations, and a sum of neutrino masses 
\begin{equation}
    \sum{m_\nu} = 0.059\ \mathrm{eV},
\end{equation}
with the neutrino masses assumed to normal-ordered. We perform a parameter fit using the COBAYA Monte-Carlo Markov Chain (MCMC) code \cite{Torrado:2020dgo} with a convergence criterion for the Gelman-Rubin $R$-parameter of $R - 1 < 0.01$. Parameters are constrained using the Planck 2018 TTTEEE power spectra \cite{Planck:2019nip} and Planck lensing \cite{Planck:2018lbu} using the NPIPE likelihood code \cite{2022MNRAS.517.4620R}, and the Atacama Cosmology Telescope (ACT) Data Release 6 lensing likelihood \cite{ACT:2023kun}. Baryon Acoustic Oscillation (BAO) constraints are from the Dark Energy Spectroscopic Instrument (DESI) DR1 data release \cite{DESI:2024mwx}. The CMB power spectrum is obtained using a modified version of the CLASS code \cite{lesgourgues2011cosmic}, which includes the $y$-distortion to the neutrino distribution function \cite{Lesgourgues_2011,Cuoco:2005qr}. We constrain parameters with and without a $y$-distortion for comparison. 

\begin{figure}
\centering
\includegraphics[width=\textwidth]{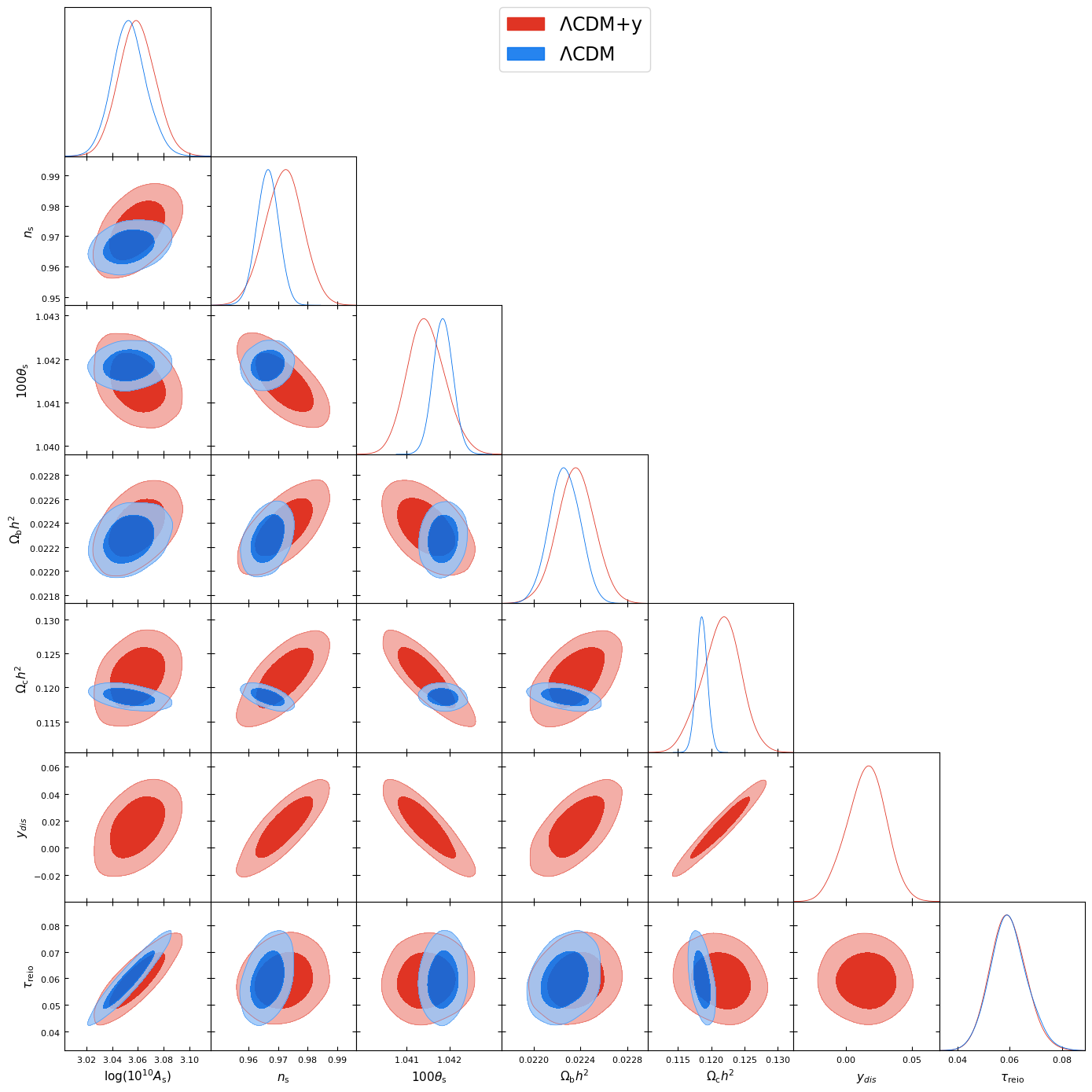}
\caption{Constraints on the the 6-parameter base $\Lambda$CDM model, and the 7-parameter model including the neutrino $y$-distortion, from Planck + Lensing + BAO.}
\label{fig:BaseParams}
\end{figure}

\begin{figure}
\centering
\includegraphics[width=0.8\textwidth]{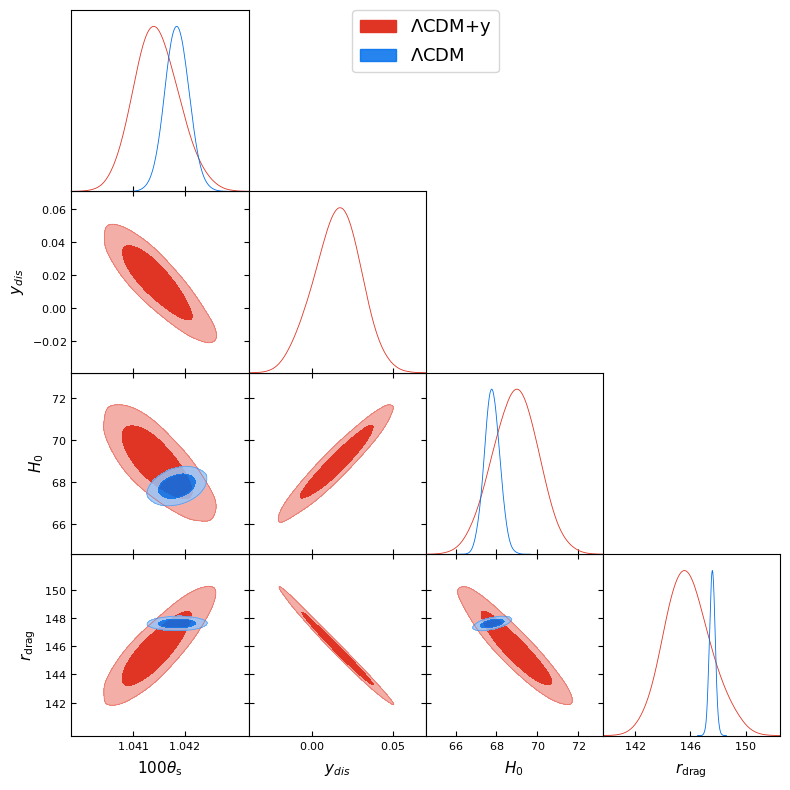}
\caption{Constraints on the Hubble constant $H_0$ and related parameters, from the 6-parameter $\Lambda$CDM model (blue), and the 7-parameter model including the neutrino $y$-distortion (red), from Planck + Lensing + BAO.}
\label{fig:H0Params}
\end{figure}

Figure \ref{fig:BaseParams} shows the constraints on the seven base parameters of the $\Lambda$CDM + $y$ model. Figure \ref{fig:H0Params} shows constraints on parameters relevant to the determination of the Hubble constant. As expected, we see a tight negative correlation between the $y$-parameter and $r_S^{\mathrm{drag}}$, so that increasing $y$ decreases the sound horizon size. The correlation between $y$ and the Hubble constant is positive: increasing $y$ increases the best-fit value for $H_0$. Constraints on the parameters individually are:
\begin{itemize}
    \item{$H_0 = 68.92 \pm 1.14\ \mathrm{km/s/MpC}$,}
    \item{$y = 0.0154 \pm 0.0147$,}
    \item{$r_S^{\mathrm{drag}} = 145.82 \pm 1.71\ \mathrm{MpC}$.}
\end{itemize}
The upper $95\%$-confidence upper bound on the Hubble constant is $H_0 < 71.12$. The constraint on $\theta_*$ becomes somewhat weaker, with a larger $y$-distortion preferring a slightly smaller $\theta_*$. 

\begin{figure}
\centering
\includegraphics[width=0.8\textwidth]{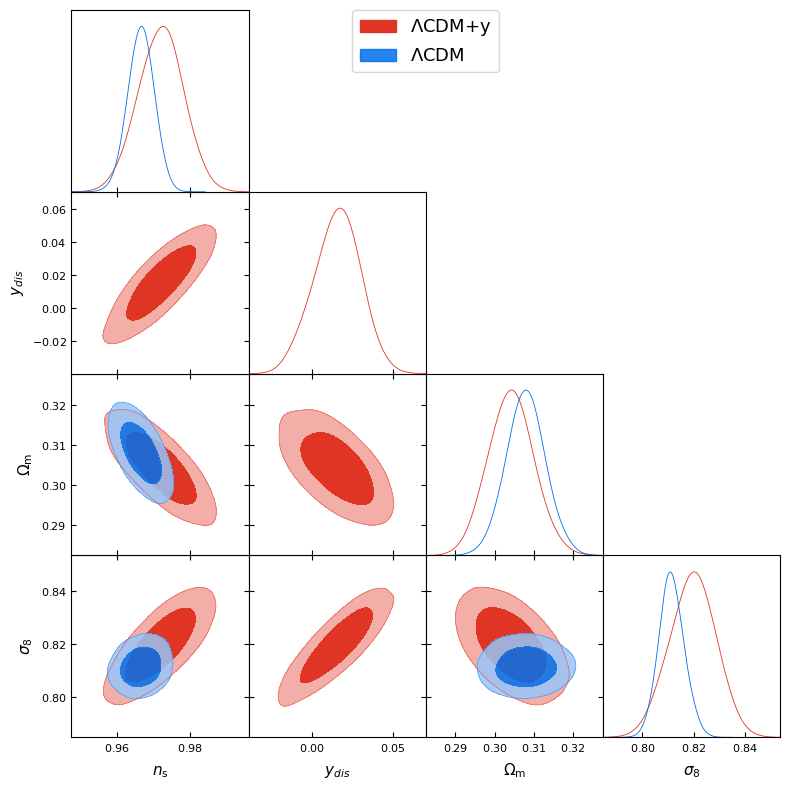}
\caption{Constraints on parameters relevant for structure formation from the 6-parameter $\Lambda$CDM model, and the 7-parameter model including the neutrino $y$-distortion, from Planck + Lensing + BAO.}
\label{fig:StructParams}
\end{figure}

Figure \ref{fig:StructParams} shows constraints on parameters relevant for structure formation. A $y$-parameter favors a somewhat lower value for $\Omega_{\mathrm{M}}$, and a slightly higher value for $\sigma_8$, although the best-fit values are not significantly shifted. The best-fit region for the spectral index $n_{\mathrm{S}}$ is shifted significantly from its value as constrained by Planck alone, with a best-fit of $n_{\mathrm{S}} = 0.9720 \pm 0.0063$, allowing for a significantly bluer spectrum, and correspondingly more power on small scales, with a larger $y$-parameter favors a higher spectral index. The fit including the $y$-distortion is $\sigma_8 = 0.8176 \pm 0.0097$, favoring a slightly higher value of the clustering parameter $\sigma_8$ than the base $\Lambda$CDM model, with a similar lower bound, so that the addition of the spectral distortion as a parameter does not serve to alleviate the tension in $\sigma_8$ between CMB and clustering measurements. (See \textit{e.g.} Chen, \textit{et al.} \cite{Chen:2022jzq}.)

\section{Discussion and Conclusions}
\label{sec:Conclusions}

In this paper, we have considered the possibility that the cosmic neutrino background, usually assumed to be thermal, has a non-thermal distribution function. For definiteness, we assume a $y$-type distortion, characterized by optically thin upscattering of neutrinos from low energy to high energy, similar to the thermal Sunyaev--Zel'dovich effect in photons. We propose no particular physical mechanism for producing such a distortion in the cosmic neutrino background, but simply treat it as physically plausible and well-defined, involving only a single additional parameter. The $y$-distortion preserves the total neutrino number, but not the total energy density in neutrinos. The dominant effect of a $y$-distortion is that the increased neutrino energy density acts in the same way as an additional (fractional) relativistic degree of freedom, $N_{\mathrm{eff}}$, with $\Delta N_{\mathrm{eff}} \simeq 4 y$. Neutrinos also contribute to structure formation, with massive nonrelativistic neutrinos behaving as a component to Cold Dark Matter, and relativistic neutrinos damping structure via free streaming. It is therefore expected that upscattering of low-energy neutrinos to high energy will decrease clustering due to the nonrelativistic component, and increase the damping due to the relativistic component. In practice, we find this effect to be negligible compared to the effect on the expansion rate. 

We place constraints on a base $\Lambda$CDM model and a one-parameter extension including the thermal $y$-parameter using the Planck 2018 TTTEEE power spectra with the NPIPE likelihood, the ACT DR6 lensing likelihood, and BAO constraints from the DESI DR1 data release. \footnote{We do \textit{not} include the ACT DR4 data for the temperature and polarization anisotropy, which is known to be in tension with Planck \cite{DiValentino:2022rdg,Handley:2020hdp,Hazra:2024nav}. In our case, the ACT data slightly favors a \textit{negative} best-fit value for the $y$-parameter.} Our main result is a $95\%$-confidence upper bound on a thermal $y$-distortion in the cosmic neutrino background of $y \leq 0.043$. We find that inclusion of a thermal $y$-distortion in the neutrino background allows for a higher value of the Hubble constant, with a best-fit of $H_0 = 68.92 \pm 1.14\ \mathrm{km/s/Mpc}$ and a $95\%$-confidence upper bound of $H_0 < 71.12$. We also find that the $y$-parameter allows for a higher spectral index for scalar perturbations, with a best-fit of $n_{\mathrm{S}} = 0.9720 \pm 0.0063$ and a $95\%$-confidence upper bound of $n_{\mathrm{S}} \leq 0.9842$. 

\section*{Acknowledgements}

This work is supported by the National Science Foundation under grant NSF-PHY-2310363. WHK thanks the Universitat de València for hospitality while a portion of this work was being completed. We thank Licia Verde for helpful conversations. 
GB is supported by the Spanish grants  CIPROM/2021/054 (Generalitat Valenciana), PID2020-113775GB-I00 (AEI/10.13039/501100011033), and by the European ITN project HIDDeN (H2020-MSCA-ITN-2019/860881-HIDDeN), 

\bibliography{Paper.bib}
\end{document}